\documentstyle[fleqn,iopfts,12pt]{ioplppt}
\jl{1}
\eqnobysec
\font\bfs=cmbxti10 scaled 1728
\font\bfss=cmbxti10 scaled 1200
\def\bq{\hbox{\bfs\symbol{113}}}
\def\sbq{\hbox{\bfss\symbol{113}}}
\begin{document}
\title{Coherent states for the \bq-deformed quantum mechanics on a circle}
\author{K Kowalski and  J Rembieli\'nski}
\address{Department of Theoretical Physics, University
of \L\'od\'z, ul.\ Pomorska 149/153,\\ 90-236 \L\'od\'z,
Poland}
\begin{abstract}
The $q$-deformed coherent states for a quantum particle on a circle 
are introduced and their properties investigated.
\end{abstract}
\section{Introduction}
In spite of the fact that the first paper devoted to the quantum
mechanics on a circle is most probably the article by Condon on the
quantum pendulum, dated 1928 [1], the coherent states for a
quantum particle on a circle have been introduced only recently
[2,3].  No wonder that there is also not any example of quantum
deformations of these states so far.  We recall that the
$q$-deformation of the standard coherent states was constructed
fifteen years ago in [4].  The need for such deformations is motivated,
among others, by the importance of the non-deformed coherent states for 
the quantum mechanics on a circle, which have been already applied for 
example in quantum gravity [5].  In this work we introduce the 
$q$-generalization of the coherent states for a quantum particle on a circle.  
We first recall the basic facts about the quantum mechanics on a circle.  
Consider a free particle on a circle $S^1$.  According to [2], the best 
candidate to represent the position of a particle on the unit circle is the 
unitary operator $U = e^{{\rm i}\hat\varphi}$ ($\hat\varphi$ hermitian) 
satisfying the following commutation rule with the hermitian angular momentum 
operator $J$:
\begin{equation}
[J,U] = U.
\end{equation}
In the Hilbert space $L^2(S^1)$ of $2\pi$-periodic functions, specified
by the scalar product
\begin{equation}
\langle f|g\rangle =\frac{1}{2\pi}\int_{0}^{2\pi}d\varphi
f^*(\varphi)g(\varphi),
\end{equation}
the operators $U$ and $J$  act as follows:
\begin{equation}
Uf(\varphi)=e^{{\rm i}\varphi}f(\varphi),
\end{equation}
and
\begin{equation}
J f(\varphi) = -{\rm i}\frac{{\rm d}}{{\rm d}\varphi}f(\varphi).
\end{equation}
Consider now the eigenvalue equation
\begin{equation}
J|j\rangle = j|j\rangle.
\end{equation}
Demanding the time-reversal invariance of
representations of (1.1) we have only two possibilities left: $j$
integer and $j$ half-integer [2].  In this work we restrict for
simplicity to the case of integer $j$.  Using (1.1) one finds that
the operators $U$ and $U^\dagger$ are the ladder operators such that
\numparts
\begin{eqnarray}
U|j\rangle &=& |j+1\rangle,\\
U^\dagger |j\rangle &=& |j-1\rangle.
\end{eqnarray}
\endnumparts
Projecting (1.6$b$) on the eigenvector $|\varphi\rangle$ of $U$
satisfying
\begin{equation}
U|\varphi\rangle=e^{{\rm i}\varphi}|\varphi\rangle
\end{equation}
we get
\begin{equation}
e_j(\varphi) := \langle \varphi|j\rangle = e^{{\rm i}j\varphi}.
\end{equation}
Of course, the vectors $e_j(\varphi)$ are the basis vectors in the
representation space $L^2(S^1)$.  We finally write down the orthogonality and
completeness conditions satisfied by the vectors $|j\rangle$ of the
form
\begin{eqnarray}
&&\langle j|j'\rangle = \delta_{jj'},\\
&&\sum_{j=-\infty}^{\infty} |j\rangle\langle j|= I.
\end{eqnarray}
\section{Coherent states for the quantum mechanics on a circle}
We now recall the construction of (non-deformed) coherent states for
a quantum particle on a circle described in [2], where the coherent 
states $|\xi\rangle$ for a quantum particle on a circle, are introduced
as the solution of the eigenvalue equation
\begin{equation}
Z|\xi\rangle = \xi|\xi\rangle,
\end{equation}
where
\begin{equation}
Z = e^{-J + \hbox{$\frac{1}{2}$}}U.
\end{equation}
The convenient parametrization of $\xi$ consistent with the form (2.2) of 
the operator $Z$ is
\begin{equation}
\xi = e^{-l + {\rm i}\alpha}.
\end{equation}
We point out that the parametrization (2.3) relies on the deformation 
of the cylinder (the phase space) specified by
\begin{equation}
x=e^{-l}\cos\alpha,\qquad y=e^{-l}\sin\alpha,\qquad z=l,
\end{equation}
and then projecting the points of the obtained surface onto the
$(x,y)$ (complex) plane. Evidently, we then identify the points of
the cylinder with the plane with extracted point $(0,0)$ (the origin).

The projection of the vectors $ |\xi\rangle$ onto the basis vectors
$ |j\rangle$ is given by
\begin{equation}
\langle j|\xi\rangle = \xi^{-j}e^{-\frac{j^2}{2}}.
\end{equation}
Making use of the parameters $l$, and $\alpha$ we can write (2.5)
in the following equivalent form:
\begin{equation}
\langle j|l,\alpha\rangle =
e^{lj-{\rm i}j\alpha}e^{-\frac{j^2}{2}},
\end{equation}
where $|l,\alpha\rangle\equiv|\xi\rangle$, with $\xi=
e^{-l + {\rm i}\alpha}$.

The coherent states are not orthogonal.  The overlap integral is [2]
\begin{equation}
\langle \xi|\eta\rangle =
\sum_{j=-\infty}^{\infty}(\xi^*\eta)^{-j}e^{-j^2} =
\theta_3\left(\frac{{\rm i}}{2\pi}\ln\xi^*\eta\Bigg\vert\frac{{\rm i}}
{\pi}\right),
\end{equation}
where $\theta_3$ is the Jacobi theta-function [6].
\section{\sbq-deformed coherent states for a particle on a circle}
We first introduce a $q$-deformation of the algebra (1.1).  The 
general deformation of the quantum mechanics on a circle was
considered in [7].  In this work we consider a $q$-deformation of
the algebra (1.1) generated only by $U$, $J_q$ and the identity
operator, of the form
\begin{equation}
qUJ_q = J_qU - U,
\end{equation}
where $U$ is a unitary operator representing the position of a 
quantum particle on a circle.  We also restrict to the case of $q>0$.  
The fact that $q$ is real follows directly from (3.1) and the assumed 
hermicity of $J_q$ and unitarity of $U$.  Now, it can be easily verified 
that the $q$-deformed angular momentum satisfying (3.1) is an element of the 
enveloping algebra of (1.1) such that
\begin{equation}
J_q = \frac{q^J-1}{q-1},
\end{equation}
while $U$ remains undeformed.  Clearly, $J_q$ acts on the basis vectors 
$|j\rangle$ as follows:
\begin{equation}
J_q |j\rangle = [j]_q |j\rangle,
\end{equation}
where $[j]_q=\frac{q^j-1}{q-1}$ is a quantum integer.  Furthermore,
using (1.4) and (3.2) we find that the action of the operator $J_q$
in $L^2(S^1)$ is of the following form:
\begin{equation}
J_qf(e^{{\rm i}\varphi}) = e^{{\rm i}\varphi}D_qf(e^{{\rm i}\varphi}),
\end{equation}
where $D_q$ designates the Jackson derivative defined by
\begin{equation}
D_qf(x) = \frac{f(qx)-f(x)}{(q-1)x},
\end{equation}
and we utilized the fact that in the discussed case of $L^2(S^1)$
spanned by the functions $e^{{\rm i}j\varphi }$, where $j$ is an integer, 
a $2\pi$-periodic function of $\varphi$ is assumed to have the Fourier 
series expansion and it can be considered as a function of 
$e^{{\rm i}\varphi}$.  Furthermore, taking into account (3.2) we get
\begin{equation}
[J_q,\hat \varphi] = -{\rm i}\ln q\,J_q - {\rm i}\frac{\ln q}{q-1},
\end{equation}
where $\hat\varphi=-i\ln U$.  We remark that the commutator (3.6) is 
nontrivially defined except of the case $q=1$ [8].

Now we are in a position to define $q$-deformed coherent states for
the quantum mechanics on a circle.  Proceeding analogously as in the
case of non-deformed coherent states discussed in the previous
section we define the $q$-deformed coherent states as a solution of
the eigenvalue equation
\begin{equation}
Z_q|\xi\rangle_q = \xi_q|\xi\rangle_q,
\end{equation}
where
\begin{equation}
Z_q := e^{{\rm i}(\hat \varphi + {\rm i}J_q)}.
\end{equation}
Using (3.6) we find after some calculation
\begin{equation}
Z_q = e^{\frac{1}{1-q}\left(\frac{1-q^{-1}}{\ln q}-1\right)}
e^{\frac{q^{-1}-1}{\ln q}J_q}U.
\end{equation}
Taking into account (3.7), (3.9), (3.3) and (1.6$b$) we get
\begin{equation}
\langle j|\xi\rangle_q = \xi_q^{-j}e^{-\frac{1}{\ln q}\frac{q^j-1}{q-1}}
e^{-\frac{j}{1-q}}.
\end{equation}
Now, the form of the operator (3.8) and (2.3) indicate the following
parametrization of a complex number $\xi_q$:
\begin{equation}
\xi_q = e^{-[l]_q+{\rm i}\alpha},
\end{equation}
where $[l]_q=\frac{q^l-1}{q-1}$, where $l$ is a real number, is
quantum $l$.  It should be noted that for $0<q<1$, $[l]_q$ is an
increasing function of $l$ and it has an upper bound $1/(1-q)$
approached in the limit $l\to+\infty$.  Consequently, (see (2.4)) in
the case of $0<q<1$, we identify the points of the $q$-deformed
classical phase space with the $(x,y)$ plane with extracted disk
$x^2+y^2\le e^{-\frac{2}{1-q}}$.  Evidently, this disk reduces to
the point $(0,0)$, i.e.\ the origin, in the limit $q\to1$ referring to
the non-deformed case.  In the case of $q>1$, $[l]_q$ is also an
increasing function and it has a lower bound $-1/(q-1)$ reached in
the limit $l\to-\infty$.  Therefore in this case we identify the
points of the deformed classical phase space with the disk
$x^2+y^2<e^{\frac{2}{q-1}}$, with extracted point $(0,0)$.  Obviously, 
in the limit $q\to1$ corresponding to the non-deformed case, the disk 
is simply the plane $(x,y)$ with extracted origin.

Using (3.11) we can write (3.10) in the form
\begin{equation}
\langle j|l,\alpha\rangle_q = e^{-{\rm i}j\alpha}e^{-\frac{1}{\ln q}
\frac{q^j-1}{q-1}}e^{-\frac{q^l}{1-q}j},
\end{equation}
where $|l,\alpha\rangle_q\equiv|\xi\rangle_q$, and $\xi_q$ referring
via (3.7) to $|\xi\rangle_q$, is given by (3.11).  Clearly, the relation
(2.6) refers to the limit $q\to1$ in (3.12).  We recall that the functions
(2.6) span the Bargmann representation [2].  The problem of finding the 
Bargmann representation in the discussed case of the $q$-deformed coherent 
states for a quantum particle on a circle is complicated and it will be
discussed in a separate work.

The coherent states are not orthogonal.  Indeed, making use of
(3.12) and (1.10) we find
\begin{equation}
{}_q\langle l,\alpha|h,\beta\rangle_q = \sum_{j=-\infty}^{\infty}
e^{{\rm i}(\alpha -\beta)j}e^{-\frac{2}{\ln q}\frac{q^j-1}{q-1}}
e^{-\frac{q^l+q^h}{1-q}j}.
\end{equation}
In the limit $q\to1$ we recover from (3.13) the following formula [2]:
\begin{equation}
\langle l,\alpha|h,\beta\rangle =
\theta_3\left(\frac{1}{2\pi}(\alpha-\beta)-\frac{l+h}{2}
\frac{{\rm i}}{\pi}\Bigg\vert\frac{{\rm i}}{\pi}\right).
\end{equation}
It follows immediately from (3.13) that the discussed coherent
states are not normalized.  Namely, we have
\begin{equation}
{}_q\langle l,\alpha|l,\alpha\rangle_q = \sum_{j=-\infty}^{\infty}
e^{-\frac{2}{\ln q}\frac{q^j-1}{q-1}}e^{-\frac{2q^l}{1-q}j}.
\end{equation}
For $q=1$ this relation reduces to
\begin{equation}
\langle l,\alpha|l,\alpha\rangle =
\sum_{j=-\infty}^{\infty}e^{2lj}e^{-j^2}=\theta_3
\left(\frac{{\rm i}l}{\pi}\Bigg\vert\frac{{\rm i}}{\pi}\right).
\end{equation}
One can easily check that the series (3.15) is convergent for
arbitrary (positive) $q$ and (finite) $l$.  Therefore, by virtue of
the Schwartz's inequality the series (3.13) is also convergent for
arbitrary $q$ and $l$.  We finally write down the formula on the
coherent states in $L^2(S^1)$ following directly from
(1.8), (1.10) and (3.12) such that
\begin{equation}
\langle \varphi|l,\alpha\rangle_q = \sum_{j=-\infty}^{\infty}
e^{{\rm i}j(\varphi-\alpha)}e^{-\frac{1}{\ln q}\frac{q^j-1}{q-1}}
e^{-\frac{q^l}{1-q}j}.
\end{equation}
In the limit $q\to1$ this formula takes the form
\begin{equation}
\langle\varphi|l,\alpha\rangle =\theta_3
\left(\frac{1}{2\pi}(\varphi-\alpha-{\rm i}l)\Bigg\vert
\frac{{\rm i}}{2\pi}\right).
\end{equation}
Notice that in view of (3.14), (3.16) and (3.18) the series from the
right-hand side of (3.13), (3.15) and (3.17), respectively, can be
regarded as $q$-deformations of the Jacobi theta functions.
\section{\sbq-deformed coherent states and the classical phase space}
In this section we discuss the parametrization (3.11) of the deformed
phase space in a more detail.  Consider the expectation value of the
deformed angular momentum operator $J_q$ given by (3.2).  On using
(1.10), (3.3) and (3.12) we arrive at the following relation:
\begin{equation}
\frac{{}_q\langle l,\alpha|J_q|l,\alpha\rangle_q}
{{}_q\langle l,\alpha|l,\alpha\rangle_q}=\frac{
\sum_{j=-\infty}^{\infty}
\frac{q^j-1}{q-1}e^{-\frac{2}{\ln q}\frac{q^j-1}{q-1}}e^{-\frac{2q^l}{1-q}j}}{
\sum_{j=-\infty}^{\infty}
e^{-\frac{2}{\ln q}\frac{q^j-1}{q-1}}e^{-\frac{2q^l}{1-q}j}}.
\end{equation}
It follows from numerical calculations that there are large regions
of the phase space parametrized by $q$ and $l$ such that
\begin{equation}
\frac{{}_q\langle l,\alpha|J_q|l,\alpha\rangle_q}
{{}_q\langle l,\alpha|l,\alpha\rangle_q}\approx [l]_q,
\end{equation}
where the approximation is very good.  For example, for $q=0.5$, and
$l\ge0.3$ the maximal relative error is of order 1 per cent.  The
fact that (4.2) is not valid for arbitrary $q$ and $l$ is not
surprising.  We only recall that in the case of the coherent states
for the quantum mechanics on a sphere [9] we have a condition $|{\bi
l}|\ge10$, where $|{\bi l}|$ is the norm of the vector ${\bi l}$ of
the classical angular momentum parametrizing the coherent states,
ensuring the (approximate) coincidence of the average of the angular 
momentum operator and ${\bi l}$.  In our opinion, the meaning of the
approximate relations like (4.2) is that the coherent states are as
close as possible to the classical phase space.  We conclude that
the parameter $[l]_q$ in (3.11) can be regarded as a deformed
version of the classical angular momentum. Let us finally recall
that in the limit $q\to1$, when (4.1) reduces to
\begin{eqnarray}
&&\frac{\langle l,\varphi|J|l,\varphi\rangle}{\langle
l,\varphi|l,\varphi\rangle}=\frac{1}{2\theta_3\left(
\frac{{\rm i}l}{\pi}\big\vert\frac{{\rm i}}{\pi}\right)}\frac{{\rm d}}{{\rm d}l}
\theta_3\left(\frac{{\rm i}l}{\pi}\Bigg\vert\frac{{\rm i}}{\pi}\right)\\\nonumber
&&{}= l - 2\pi\sin
(2l\pi)\sum_{n=1}^{\infty}\frac{e^{-\pi^2(2n-1)}}{(1+e^{-\pi
^2(2n-1)}e^{2{\rm i}l\pi})(1+e^{-\pi^2(2n-1)}e^{-2{\rm i}l\pi})}\,,
\end{eqnarray}
we have a perfect approximation of the classical phase space for
arbitrary $l$ [2]. Namely, the maximal error of (4.2) is of order $0.1$ 
per cent, and we have the {\em exact\/} equality in the case with $l$ 
integer or half-integer.

We now examine the role of the parameter $\alpha$ in the
parametrization (3.11). Taking into account (1.10), (1.6$a$) and
(3.12) we find
\begin{equation}
\frac{{}_q\langle l,\alpha|U|l,\alpha\rangle_q}
{{}_q\langle l,\alpha|l,\alpha\rangle_q}=e^{{\rm i}\alpha}\frac{
e^{\frac{q^l-1}{q-1}-\frac{1}{\ln q}-\frac{1}{1-q}}
\sum_{j=-\infty}^{\infty}
e^{-\frac{1+q}{\ln q}\frac{q^j-1}{q-1}}e^{-\frac{2q^l}{1-q}j}}{
\sum_{j=-\infty}^{\infty}
e^{-\frac{2}{\ln q}\frac{q^j-1}{q-1}}e^{-\frac{2q^l}{1-q}j}}.
\end{equation}
In the limit $q\to$ this formula takes the form
\begin{equation}
\frac{\langle l,\alpha|U|l,\alpha\rangle}{\langle
l,\alpha|l,\alpha\rangle} =
e^{-\frac{1}{4}}e^{{\rm i}\alpha}\,\frac{\theta_2\left(\frac{{\rm i}l}{
\pi}\big\vert\frac{{\rm i}}{\pi}\right)}{\theta_3\left(\frac{{\rm i}l}{
\pi}\big\vert\frac{{\rm i}}{\pi}\right)}=e^{-\frac{1}{4}}e^{{\rm i}\alpha}\,
\frac{\theta_3\left(l+\hbox{$\frac{1}{2}$}\big\vert{\rm i}\pi\right)}
{\theta_3\left(l\big\vert{\rm i}\pi\right)}.
\end{equation}
Proceeding analogously as in [2] we define the relative expectation value
\begin{equation}
{}_q\langle\!\langle U\rangle\!\rangle_{(l,\alpha)}=\frac{{}_q\langle U
\rangle_{(l,\alpha)}}{{}_q\langle U\rangle_{(l,0)}},
\end{equation}
where ${}_q\langle U\rangle_{(l,\alpha)}={}_q\langle l,\alpha|U|l,
\alpha\rangle_q/{}_q\langle l,\alpha|l,\alpha\rangle_q$.  Hence
\begin{equation}
{}_q\langle\!\langle U\rangle\!\rangle_{(l,\alpha)}=e^{{\rm i}\alpha}.
\end{equation}
Therefore, the parameter $\alpha$ labelling the coherent states can
be identified with the classical angle.

We now study the distribution of eigenvectors $|j\rangle$'s of
the operator $J_q$ in the normalized coherent state.  We recall that
in the non-deformed case this is the distribution of energies of a
quantum particle moving freely in a (unit) circle [2].  The computer
simulations indicate that the function (see (3.12) and (3.15))
\begin{equation}
p_{l,q}(j) = \frac{|\langle j|l,\alpha\rangle_q|^2}{{}_q\langle
l,\alpha|l,\alpha\rangle_q} = \frac{
e^{-\frac{2}{\ln q}\frac{q^j-1}{q-1}}e^{-\frac{2q^l}{1-q}j}}
{\sum_{j=-\infty}^{\infty}
e^{-\frac{2}{\ln q}\frac{q^j-1}{q-1}}e^{-\frac{2q^l}{1-q}j}},
\end{equation}
which gives the probability of finding the system in the state 
$|j\rangle$ when the system is in the normalized coherent state 
$|l,\alpha\rangle_q/\sqrt{{}_q\langle l,\alpha|l,\alpha\rangle_q}$,
has the behaviour the same as in the non-deformed case, that is
$p_{l,q}(x)$ is peaked at $x=l$.  However, in opposition to the
distribution referring to the non-deformed coherent states, when $q=1$, 
which is the ``discrete'' Gaussian distribution of the form [2]
\begin{equation}
p_l(j) = \frac{|\langle j|l,\alpha\rangle|^2}{\langle
l,\alpha|l,\alpha\rangle} = \frac{e^{2lj}e^{-j^2}}{
\theta_3\left(\frac{{\rm i}l}{\pi}\big\vert\frac{{\rm i}}{\pi}\right)}
\approx \frac{e^{-(j-l)^2}}{\sqrt{\pi}},
\end{equation}
where the approximation is very good, the distribution (4.8) is
asymmetrical one (see figure 1).

We finally discuss the probability density
$p_{(l,\alpha),q}(\varphi)$ for the coordinates in the normalized
coherent state $|l,\alpha\rangle_q/\sqrt{{}_q\langle l,\alpha|l,
\alpha\rangle_q}$, of the form (see (3.17))
\begin{equation}
p_{(l,\alpha),q}(\varphi) = \frac{|\langle \varphi|l,\alpha\rangle_q|^2}
{{}_q\langle l,\alpha|l,\alpha\rangle_q} = \frac{\big\vert
\sum_{j=-\infty}^{\infty}
e^{{\rm i}j(\varphi-\alpha)}e^{-\frac{1}{\ln q}\frac{q^j-1}{q-1}}
e^{-\frac{q^l}{1-q}j}\big\vert^2}
{\sum_{j=-\infty}^{\infty}
e^{-\frac{2}{\ln q}\frac{q^j-1}{q-1}}e^{-\frac{2q^l}{1-q}j}}.
\end{equation}
As with the non-deformed case (see (3.18) and (3.16)) the function
$p_{(l,\alpha),q}(\varphi)$ is peaked at $\varphi=\alpha$ (see
figure 2).  Therefore, the parameter $\alpha$ in (3.11) can be
really regarded as the classical angle.
\section{A generalization of the \sbq-deformed coherent states}
We finally study a generalization of the $q$-deformed coherent states
discussed above arising from taking into consideration the so called
``squeezed states'' introduced in [10].  These states amount a
generalization of the non-deformed coherent states for the quantum
mechanics on a circle introduced in [2] and can be regarded as a
version of the coherent states on a circle introduced in [3].
Namely, they can be defined as a solution of the eigenvalue equation
generalizing (2.1) such that [10]
\begin{equation}
Z(s)|\xi\rangle_s = \xi|\xi\rangle_s,
\end{equation}
where
\begin{equation}
Z(s) = e^{-s\left(J - \hbox{$\frac{1}{2}$}\right)}U,
\end{equation}
and $s>0$ is a real parameter.  Clearly, the case of $s=1$
corresponds to the coherent states discussed in section 2.  An
attempt to provide a physical interpretation of this dimensionless 
parameter was made in [3] and [11], where it is suggested that it
controls the ratio of spatial width of the coherent states to the
length of the circle.  The counterparts of the relations (2.5) and
(2.7) are of the form
\begin{eqnarray}
\langle j|\xi\rangle_s &=& \xi^{-j}e^{-\frac{sj^2}{2}},\\
{}_s\langle \xi|\eta\rangle_s &=&
\sum_{j=-\infty}^{\infty}(\xi^*\eta)^{-j}e^{-sj^2} =
\theta_3\left(\frac{{\rm i}}{2\pi}\ln\xi^*\eta\Bigg\vert\frac{{\rm is}}
{\pi}\right).
\end{eqnarray}
Now, we define the generalized $q$-deformed coherent states for the
quantum mechanics on a circle as the solution of the eigenvalue
equation
\begin{equation}
Z_q(s)|\xi\rangle_{s,q} = \xi_q|\xi\rangle_{s,q},
\end{equation}
where
\begin{equation}
Z_q(s) := e^{{\rm i}(\hat \varphi + {\rm i}sJ_q)}
= e^{\frac{s}{1-q}\left(\frac{1-q^{-1}}{\ln q}-1\right)}
e^{\frac{s(q^{-1}-1)}{\ln q}J_q}U.
\end{equation}
Of course, the states $|\xi\rangle_q$ given by (3.7) refer to the
case with $s=1$.  Using (5.5) and (5.6) we easily obtain the
following generalizations of the relations (3.10), (3.13) and
(3.15), respectively:
\begin{eqnarray}
&&\langle j|\xi\rangle_{s,q} = \xi_q^{-j}e^{-\frac{s}{\ln q}\frac{q^j-1}{q-1}}
e^{-\frac{sj}{1-q}},\\
&&{}_{s,q}\langle l,\alpha|h,\beta\rangle_{s,q} = \sum_{j=-\infty}^{\infty}
e^{{\rm i}(\alpha -\beta)j}e^{-\frac{2s}{\ln q}\frac{q^j-1}{q-1}}
e^{-\frac{q^l+q^h}{1-q}j}e^{\frac{2(1-s)}{1-q}j},\\
&&{}_{s,q}\langle l,\alpha|l,\alpha\rangle_{s,q} = \sum_{j=-\infty}^{\infty}
e^{-\frac{2s}{\ln q}\frac{q^j-1}{q-1}}e^{-\frac{2q^l}{1-q}j}
e^{\frac{2(1-s)}{1-q}j}.
\end{eqnarray}
Applying the d'Alambert ratio test we find that the series (5.9) is
not convergent for arbitrary (positive) $s$.  Namely, it follows
that it is convergent if $q^l>1-s$, and divergent if $q^l<1-s$.
Notice that these conditions seem to distinguish the case $s=1$
discussed earlier, because only if $s=1$ the series (5.9) is
convergent for arbitrary $l$.
\section{Discussion}
In this work we have introduced the $q$-deformed coherent states for
the quantum mechanics on a circle.  The correctness of the
construction is confirmed by the quasi-classical character of the
coherent states manifested for example by the behaviour of the
expectation values of the deformed angular momentum operator.  It is
worthwhile to recall that the non-deformed coherent states specified by  
(2.1) as well as the coherent states of a quantum particle on a sphere
introduced by us in [9] are concrete realization of the general 
mathematical scheme of construction of the Bargmann spaces introduced
in the recent papers [11], [12] and [13].  Thus, bearing in mind the
observations presented herein, an interesting problem naturally arises
of finding deformations of coherent states for the quantum mechanics on
a sphere.  It should be noted that in view of the observations of Dimakis 
and M\"uller-Hoissen [14] one can relate by the suitable change of 
variables, the $q$-deformation of the quantum mechanics on a circle 
described by the Jackson derivative (3.5) with a discrete quantum 
mechanics on a lattice.  We finally remark that the results of this
paper would be of importance in the theory of special functions.  We
only recall that the formulae (3.13), (3.15) and (3.17) describe a quantum
deformation of the Jacobi theta functions.
\ack
This paper has been supported by the Polish Ministry of Scientific 
Research and Information Technology under the grant 
No PBZ-MIN-008/P03/2003.

\Figures
\begin{figure}
\caption{The plot of $p_{l,q}(j)$ (see (4.8)), where $q=0.5$
and $l=2$. The maximum is reached at $j_{{\rm max}}=l$.}
\label{fig1}
\caption{The probability density $p_{(l,\alpha),q}(\varphi)$ given by
(4.10), where $q=0.5$, $l=1$ and $\alpha=\pi$.  The probability
density is peaked at $\varphi=\alpha$.}
\label{fig2}
\end{figure}
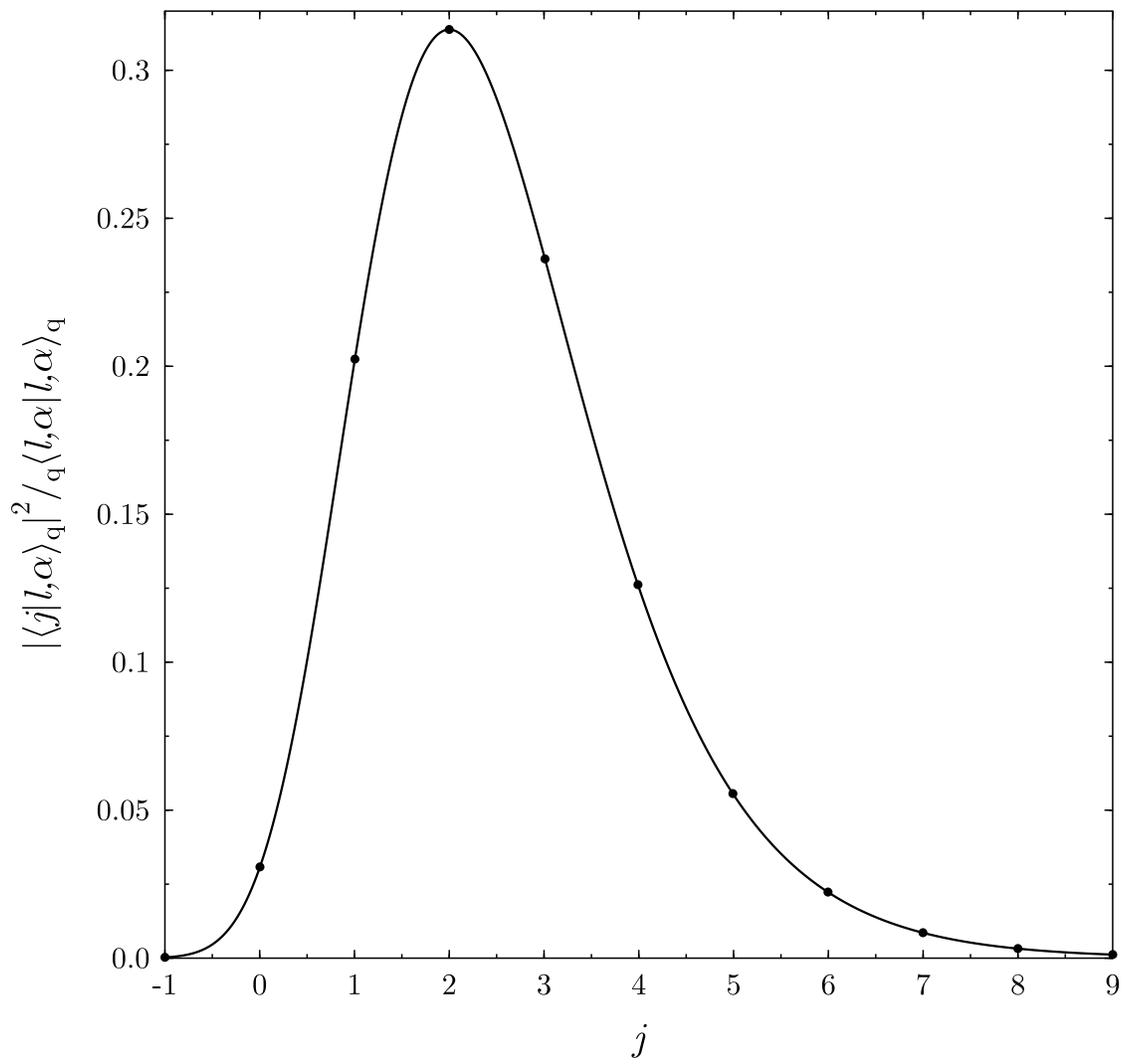
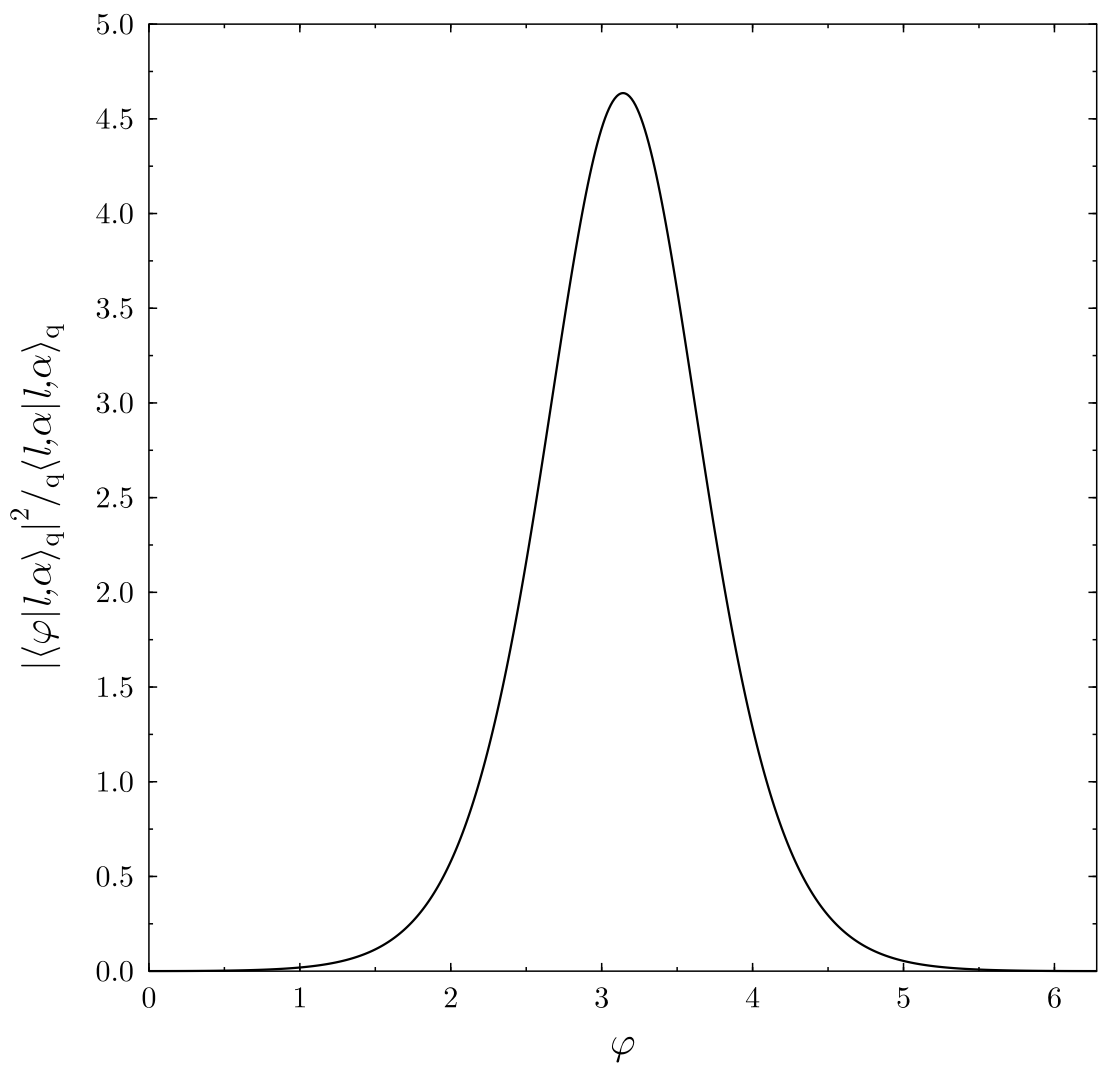
\end{document}